\documentstyle[12pt,a4,fleqn]{article}

\title{Emission of gamma rays shifted from resonant absorption by
electron-nuclear double transitions in $^{151}$Eu$^{2+}$:CaF$_2$}
\author{Silviu Olariu\\
Institute of Physics and Nuclear Engineering, Heavy-Ion Physics Department\\
76900 Magurele, P.O. Box MG-6, Bucharest, Romania\\
\and
James J. Carroll\\
Department of Physics and Astronomy, Youngstown State University\\
Youngstown, Ohio 44555\\
\and
Carl B. Collins\\
Center for Quantum Electronics, University of Texas at Dallas\\
P.O.Box 830688, Richardson, Texas 75083-0688\\}

\begin{document}
\date{}
\maketitle
\date{}

\abstract
We show that the emission of a gamma-ray photon by a nucleus can be influenced
by a microwave magnetic field acting on the atomic electrons. We study 
theoretically these
electron-nuclear double transitions (ENDTs) for $^{151}$Eu nuclei in a CaF$_2$
lattice at low temperature, in the presence of a static magnetic field and
of a microwave magnetic field. The ENDTs acquire a significant intensity for
certain resonance frequencies. The ENDTs are of interest  for the 
identification of the position of the lines in complex M\"{o}ssbauer spectra.\\

PACS numbers: 23.20.Lv, 76.70.Dx, 76.80.+y\\
\endabstract

\setlength{\parindent}{0cm}
\newpage
The possibility of influencing gamma-ray emission with applied oscillatory
electromagnetic fields was first analyzed by Hack and Hamermesh. \cite{1} 
In the present work we show that the emission of a gamma-ray photon by a 
nucleus can
be influenced by a microwave magnetic field acting on the atomic electrons.
We consider a two-photon process in which a microwave photon is absorbed
by the electrons simultaneously with the emission of a gamma-ray photon by the
nucleus. We study theoretically these electron-nuclear double transitions 
(ENDTs) for
$^{151}$Eu nuclei in a CaF$_2$ lattice at low temperature, in the presence of
a static magnetic field and under the action of a microwave magnetic field.
The $^{151}$Eu nucleus is supposed to be initially in the 21.54 keV excited
state and the atomic electrons are in the ground state. We show that, under
the action of the microwave magnetic field, the coupled electron-nuclear 
system can make a transition from the ground atomic electron Zeeman level 
to a higher electron level while at the same time the nucleus makes a 
transition from the 21.54 keV state to the ground nuclear state emitting a
gamma-ray photon. We show that the energy of the gamma-ray photon emitted by
ENDTs is shifted away from the parent gamma-ray line, and obtain for the
ENDT lines a quadratic dependence on the microwave magnetic field and an
inversely quadratic dependence on the detuning between the position of the
virtual and intermediate states. The ENDT lines acquire a significant
intensity for certain resonance frequencies. The ENDTs are of interest for 
the identification of the position of the lines in complex 
M\"{o}ssbauer spectra.\\

We shall analyze in this work the 21.54 keV, M1 transition of $^{151}$Eu nuclei
embedded in a CaF$_2$ lattice. The parity and the nuclear magnetic moment 
of the ground
state are 5/2+, 3.474 $\mu_N$, and the spin, the parity, the nuclear magnetic
moment and the half-life of the 21.54 keV state are respectively 7/2+,
2.591 $\mu_N$ and 9.4 ns, where the nuclear magneton is 
$\mu_N=5.05\cdot 10^{-27}$ JT$^{-1}$. \cite{9} The 18 hyperfine components
of the 21.54 keV transition give rise, by overlapping, to an 8-line
M\"{o}ssbauer spectrum.\cite{10},\cite{11} Europium appears in a CaF$_2$ 
lattice as the divalent ion Eu$^{2+}$, having the ground state $^8S_{7/2}$.
The electron energy levels of the Eu$^{2+}$ ion in the cubic field of the
CaF$_2$ lattice have been studied by Lacroix \cite{12} and experimentally
observed with the method of electron paramagnetic resonance by Ryter \cite{13}
and Baker et al. \cite{14} The results of ENDOR observations on 
$^{151}$Eu$^{2+}$ in a CaF$_2$ lattice have been reported by Baker and
Williams. \cite{15}\\

We shall write the Hamiltonian of the $^{151}$Eu$^{2+}$ system in the CaF$_2$
lattice as the sum
$H=H_e+H_n+H_{en}+V_{mw}(t)+V_\gamma (t) \:,$
where $H_e$ is the electron Hamiltonian, $H_n$ is the nuclear Hamiltonian,
$H_{en}$ describes the hyperfine interaction, $V_{mw}(t)$ describes the
interaction of the microwave field with the atomic electrons and $V_\gamma(t)$
describes the interaction of the gamma rays with the nucleus.\\

We shall assume that the x, y, z axes are oriented along the four-fold axes
of the CaF$_2$ crystal, with the origin on a Eu$^{2+}$ ion. If the static
magnetic field $B$ is applied along the z-axis, and if we denote the 
projections of the electron angular momentum along the z-axis by $M$, where
$M=\pm7/2, \pm5/2, \pm3/2, \pm1/2$, it can be shown \cite{12},\cite{16}
that the eigenvalues $E_M$ of the electron Hamiltonian $H_e$ are
\begin{equation}
E_{\pm7/2}=\pm\frac{3}{2}g\mu_B B+8b_4-2b_6\pm\left(\left(\pm2g\mu_B
B-b_4+3b_6\right)^2+35\left(b_4-3b_6\right)^2\right)^{1/2} \:,
\end{equation}
\begin{equation}
E_{\pm5/2}=\pm\frac{1}{2}g\mu_B B-8b_4+2b_6\pm\left(\left(\pm2g\mu_B
B-5b_4-7b_6\right)^2+3\left(5b_4+7b_6\right)^2\right)^{1/2} \:,
\end{equation}
\begin{equation}
E_{\pm3/2}=\mp\frac{1}{2}g\mu_B B-8b_4+2b_6\pm\left(\left(\pm2g\mu_B
B+5b_4+7b_6\right)^2+3\left(5b_4+7b_6\right)^2\right)^{1/2} \:,
\end{equation}
\begin{equation}
E_{\pm1/2}=\mp\frac{3}{2}g\mu_B B+8b_4-2b_6\pm\left(\left(\pm2g\mu_B
B+b_4-3b_6\right)^2+35\left(b_4-3b_6\right)^2\right)^{1/2} \:,
\end{equation}
where the Bohr magneton is $\mu_B=9.274\cdot 10^{-24}$ JT$^{-1}$, and
$g$=1.9926, $b_4$=-176.12 MHz, $b_6$=0.78 MHz. \cite{15},\cite{17}
Then the eigenvalues of the time-independent part $H_e+H_n+H_{en}$ of the
Hamiltonian $H$ are \cite{12}
\begin{eqnarray}
\lefteqn{E_{Mm_{l}}^{(l)}=E_M+A_lMm_l+\frac{A_l^2\left(7/2+M\right)
\left(9/2-M\right)\left(5/2-m_l\right)\left(7/2+m_l\right)}{4\left(E_M-E_{M-1}
\right)}}\nonumber \\
 & & -\frac{A_l^2\left(7/2-M\right)\left(9/2+M\right)\left(5/2+m_l\right)
\left(7/2-m_l\right)}{4\left(E_{M+1}-E_M\right)}-g_l\mu_B B m_l \:,
\end{eqnarray}
\begin{eqnarray}
\lefteqn{E_{Mm_{u}}^{(u)}=E_0+E_M+A_uMm_u+\frac{A_u^2\left(7/2+M\right)
\left(9/2-M\right)\left(7/2-m_u\right)\left(9/2+m_u\right)}{4\left(E_M-E_{M-1}
\right)}}\nonumber \\
 & & -\frac{A_u^2\left(7/2-M\right)\left(9/2+M\right)\left(7/2+m_u\right)
\left(9/2-m_u\right)}{4\left(E_{M+1}-E_M\right)}-g_u\mu_B B m_u \:,
\end{eqnarray}
In eqs. (5),(6), $E_0$=21.54 keV is the energy of the unsplit excited nuclear 
state, $m_l=\pm5/2, \pm3/2, \pm1/2$ are the projections on the z-axis of the
angular momentum of the ground state $l$ of the nucleus, and $m_u=\pm7/2, 
\pm5/2, \pm3/2, \pm1/2$ are the projections on the z-axis of the angular
momentum of the excited state $u$ of the nucleus. The ground state constants
are \cite{15},\cite{17} $A_l/(2\pi\hbar)$=-102.9069 MHz, 
$g_l$=7.4968$\cdot 10^{-4} $. The constants for the upper state $u$ can
be calculated from the relation $A_u=\mu_uI_l A_l/(\mu_l I_u), 
g_u=\mu_uI_l g_l/(\mu_l I_u)$, where $\mu_l, \mu_u$ are respectively the 
magnetic moments for the ground and excited states $l,u, I_l=5/2, I_u=7/2$,
and $\mu_uI_l/(\mu_l I_u)$=0.5327, so that $A_u/(2\pi\hbar)$=-54.8185 MHz,
$g_u$=3.9936$\cdot 10^{-4}$. The electron-nuclear states of interest for the
present work are represented in fig. 1.\\

We are now in a position to analyze the two-photon process by which a gamma-ray
photon is emitted by the nucleus simultaneously with the absorption of a
microwave photon by the atomic electrons. As already discussed, the states of 
the problem are $|M,lm_l\rangle  $ and $|M,um_u\rangle $. In the initial state 
$|-7/2,um_u\rangle $
of the two-photon process the Eu$^{2+}$ ion is in the $M$=-7/2 state and the
$^{151}$Eu nucleus is in the excited state $u$ of angular momentum $m_u$, and
in the final state $|-5/2,lm_l\rangle $ of the two-photon process the 
Eu$^{2+}$ ion 
is in the $M$=-5/2 state and the $^{151}$Eu nucleus reaches the ground state 
$l$ of angular momentum $m_l$, where $m_u-m_l=0, \pm1$. The intermediate states
of the two-photon process are $|-5/2,um_u\rangle $ and $|-7/2,lm_l\rangle $. 
We can calculate
the ENDT transition amplitude 
$c_2=\langle -5/2,lm_l|{\rm ENDT}|-7/2,um_u\rangle $ with the 
aid of
conventional second-order perturbation theory, and the single-photon gamma-ray
transition amplitude as $c_1=\langle -7/2,lm_l|V_\gamma|-7/2,um_u\rangle $, 
where the 
operator $V_\gamma$ describes the interaction of the gamma-ray photon with the
nucleus. In this way it can be shown that the relative intensity of the
ENDT lines $|c_2|^2/|c_1|^2$ increases quadratically with the applied microwave
magnetic field $B_{mw}$, and decreases with the square of the detuning between
the virtual states and the corresponding intermediate states. It turns out that
significant intensities for the ENDT lines can be obtained when the energy of
the microwave photon is resonant with the energy of the transition 
$|-7/2,um_u\rangle \rightarrow |-5/2,um_u\rangle $, so that
$\hbar\omega_{mw}=E_{-5/2,m_u}^{(u)}-E_{-7/2,m_u}^{(u)}\:.$
The gamma-ray photons emitted in the ENDT transition have the energy
$\hbar\omega_\gamma=E_{-5/2,m_u}^{(u)}-E_{-5/2,m_l}^{(l)}\:,$
and the positions of the single-photon gamma-ray lines are given by
$\hbar\omega_\gamma^{(0)}=E_{-7/2,m_u}^{(u)}-E_{-7/2,m_l}^{(l)}\:,$\\

It can be shown by conventional second-order perturbation theory that the ratio
$R=|c_2^{(r)}|^2/|c_1|^2$ of the square of the resonant ENDT  
amplitude
$c_2^{(r)}=\langle -5/2,lm_l|{\rm ENDT}_r|-7/2,um_u\rangle $ to the square of
the  unperturbed
single-photon amplitude $c_1=\langle -7/2,lm_l|V_\gamma|-7/2,um_u\rangle $ is
\begin{equation}
R=\frac{7g^2\mu_B^2B_{mw}^2}{4\hbar^2{\it \Gamma}^2}\:,
\end{equation}
being independent of $m_u,m_l$. We consider that the width ${\it \Gamma}$ is 
determined by the half-life $t_{1/2}$=9.4 ns of the 21.54 keV state of the
$^{151}$Eu nucleus as ${\it \Gamma}=\ln 2/t_{1/2}$. For a resonant microwave 
field of amplitude $B_{mw}=10^{-4}$ T, the relative intensity of the ENDT line 
is $R$=0.098.
The relative variation of the recoil-free fraction for the gamma-ray
transitions between the substates shown in 
fig. 1 is of the order of $3E_0 \hbar\omega_{mw}/(2m_{\rm Eu}k\theta_D)$ and is
negligible, where $m_{\rm Eu}$ is the mass of the europium nucleus, 
$\theta_D$ is the Debye temperature of the lattice and $k$ the Boltzmann 
constant. In the derivation of eq. (7) it has been implicitly assumed that
the half-life of the nuclear excited state is much shorter than the
spin-lattice relaxation time, an assumption which is valid for the 9.4 ns 
half-life of the 21.54 keV state of $^{151}$Eu. \cite{17b}\\

In fig. 2(a)  we have represented the ENDT spectrum for a static magnetic field
$B$=2 T, which gives $(E_{-7/2}-E_{-5/2})/(2\pi\hbar)$=59.291 GHz, for a 
sample temperature $T$=1 K. The resonance microwave frequency
for the ENDT  $|-7/2,u7/2\rangle \rightarrow |-5/2,u7/2\rangle 
\rightarrow |-5/2,l5/2\rangle $
shown in fig. 1 is $\omega_{mw}/(2\pi)=59.098$ GHz. The
ENDT line is situated at $(E_\gamma-E_0)/(2\pi\hbar)$=-148 MHz. The direction
of observation is perpendicular to the static magnetic field $B$. Assuming a
Boltzmann distribution for the electron population, the fractional electron
population is 0.941 in the $M$=-7/2 state and 0.054 in the -5/2 state. The
FWHM of the lines was taken 23.5 MHz. The intensity $|c_1|^2$ of the 
single-photon lines is proportional to $|(\mu_+)_{lm_l,um_u}|^2+
|(\mu_-)_{lm_l,um_u}|^2+|(\mu_z)_{lm_l,um_u}|^2$. In fig. 2(b) we have 
represented the ENDT spectrum for a static magnetic field
$B$=0.2342 T, which gives $(E_{-5/2}-E_{-7/2})/(2\pi\hbar)$=10 GHz, and at a 
sample
temperature $T$=1 K, for the same ENDT transition and direction of observation
as before. The resonance frequency is 9.797 GHz, and the ENDT line is situated
at $(E_\gamma-E_0)/(2\pi\hbar)$=-142 MHz. The fractional electron population is
0.31 in the $M$=-7/2 state and 0.19 in the $M$=-5/2 state. The relative 
intensity of the ENDT line, being proportional to the difference of the 
electron populations of the $M$=-7/2 and $M$=-5/2 levels, is thus smaller at
10 GHz than at 59 GHz, but the ENDT contribution can still be clearly seen
in the spectrum.\\

The process by which the $^{151}$Eu$^{2+}$ system, initially in the 
$|-7/2,u7/2\rangle $ state, absorbs a microwave photon up to the $|-5/2,u7/2
\rangle $
intermediate state before emitting a gamma-ray photon to reach the final
$|-5/2,l5/2\rangle $ state is labelled in fig. 1 as direct ENDT. The cross 
section
for the stimulated emission of a gamma-ray photon via this ENDT is denoted
by $\sigma$. There is also an ENDT when the $^{151}$Eu$^{2+}$ system, initially
in the $|-5/2,l5/2\rangle $ state, absorbs a gamma-ray photon up to the 
intermediate
$|-5/2,u7/2\rangle $ state and then emits a microwave photon to reach the final
$|-7/2,u7/2\rangle $ state. The process is displayed in fig. 1 as the inverse 
ENDT.
The cross section for the absorption of a gamma-ray photon via the inverse
ENDT will have the same value $\sigma$ as the direct ENDT.\\

Absorption ENDTs are also possible when 
the $^{151}$Eu$^{2+}$ system, initially in the $|-7/2,lm_l\rangle $ state, 
absorbs
a microwave photon up to the intermediate state $|-5/2,lm_l\rangle $, and then
absorbs a gamma-ray photon to reach the final $|-5/2,um_u\rangle $ state.
For a resonant ENDT in absorption, the energy of the absorbed gamma-ray photon
continues to be given by 
$\hbar\omega_\gamma=E_{-5/2,m_u}^{(u)}-E_{-5/2,m_l}^{(l)}\:,$, and the 
relative 
intensity of the absorbed
gamma ray by eq. (7), but the microwave resonance frequency is, for a direct
ENDT in absorption,
$\hbar\omega_{mw}^{(a)}=E_{-5/2,m_l}^{(l)}-E_{-7/2,m_l}^{(l)}\:.$
The resonance microwave frequency for the direct ENDT in absorption
$|-7/2,l5/2\rangle \rightarrow|-5/2,l5/2\rangle \rightarrow|-5/2,u7/2
\rangle $ is 
$\hbar\omega_{mw}^{(a)}/(2\pi)$=59.032 GHz for $B$=2 T, and 
$\hbar\omega_{mw}^{(a)}/(2\pi)$=9.717 GHz for $B$=0.2342 T. The diference
$\delta=(\omega_{mw}-\omega_{mw}^{(a)})/(2\pi) $ is $\delta$=66 MHz for 
$B$=2 T, and $\delta$=79 MHz for $B$=0.2342 T.\\

The ENDTs described in this work may have 
applications for the research on the amplification of gamma rays without
inversion of nuclear population. \cite{18}
Another interesting application of ENDTs would be the observation of gamma-ray 
{\it lines}
in situations when the conventional M\"{o}ssbauer spectrum is unresolved, or
partially resolved.  This case is exemplified by the spectrum of fig. 2(b), 
where there are many overlapping lines arising from the electron states of 
various $M$ values. If we record M\"{o}ssbauer spectra for a series of applied
microwave frequencies from a range where we expect to have resonances at the
value of the applied static magnetic field, the difference between these
spectra and a reference spectrum corresponding to a non-resonant microwave
frequency will show the ENDT contribution, as illustrated in the lower part
of fig. 2(b). For a particular resonance frequency, this contribution consists 
of pairs of peaks and dips, the positions of which coincide with the position
of the lines in the $M$=-7/2 and respectively $M$=-5/2 M\"{o}ssbauer spectra
originating from {\it one} initial nuclear sublevel. Therefore, the ENDT
peak/dip spectrum is much simpler than the full M\"{o}ssbauer spectrum, where
we have lines from {\it all} values of $M$. Thus, in the ENDT approach, 
the full M\"{o}ssbauer spectrum is decomposed in a series of partial 
M\"{o}ssbauer spectra obtained for the microwave resonance frequencies of the
system.\\

ACKNOWLEDGMENT\\

The work of one of the authors (S.O.) has been supported by a 
research grant of the Romanian Ministry of Research and Technology.\\

\newpage
FIGURE CAPTIONS\\

Fig. 1. Energy levels of the electron-nuclear system $^{151}$Eu$^{2+}$:CaF$_2$
relevant for the present work. For a static magnetic field $B$=2 T we have
$(E_{-5/2}-E_{-7/2})/(2\pi\hbar)$=59.291 GHz, $\omega_{mw}/(2\pi)$=59.098 GHz, 
$\omega_{mw}^{(a)}/(2\pi)$=59.032 GHz. For a static magnetic field $B$=0.2342 T
we have $(E_{-5/2}-E_{-7/2})/(2\pi\hbar)$=10 GHz, $\omega_{mw}/(2\pi)$=9.797 
Ghz, $\omega_{mw}^{(a)}/(2\pi)$=9.717 GHz. $E_0$=21.54 keV is the energy of the
excited nuclear state of $^{151}$Eu.\\

Fig. 2. Unperturbed M\"{o}ssbauer spectrum of $^{151}$Eu$^{2+}$:CaF$_2$ (upper
curve) and ENDT contribution (lower curve) at  $T$= 1 K, for  (a) $B$=2 T, 
$B_{mw}=10^{-4}$ T,  $\omega_{mw}/(2\pi)$=59.098 GHz, and 
(b) $B$=0.2342 T, $B_{mw}=10^{-4}$ T, $\omega_{mw}/(2\pi)$=10 GHz.
The FWHM of the lines was taken 23.5 MHz.\\


\newpage
\begin{thebibliography}{17}
\bibitem{1} M. N. HACK and M. HAMERMESH, {\it Nuovo Cimento}, {\bf XIX}  
(1961) 546.    
\bibitem{9} C. M. LEDERER and V. S. SHIRLEY, eds. {\it Table of Isotopes}, 
Seventh Edition (Wiley, New York, 1978).
\bibitem{10} P. H. BARRETT and D. A. SHIRLEY, {\it Phys. Rev.}, {\bf 131}
(1963)  123.
\bibitem{11} G. CRECELIUS, {\it Z. Physik}, {\bf 256} (1972)  155.
\bibitem{12} R. LACROIX, {\it Helv. Phys. Acta}, {\bf XXX} (1957)  374.
\bibitem{13} CH. RYTER, {\it Helv. Phys. Acta}, {\bf XXX} (1957)  353.
\bibitem{14} J. M. BAKER, B. BLEANEY and W. HAYES, {\it Proc. Roy. Soc. A}, 
{\bf 247} (1958)  141.
\bibitem{15} J. M. BAKER and F. I. B. WILLIAMS, {\it Proc. Roy. Soc A}, 
{\bf 267} (1962) 283.
\bibitem{16} A. ABRAGAM and B. BLEANEY, {\it Electron Paramagnetic Resonance of
Transition Ion}s (Clarendon, Oxford, 1970), pp. 149, 863.
\bibitem{17} ibid., pp. 252, 335.
\bibitem{17b} ibid., p. 573.
\bibitem{18} S. OLARIU, J. J. CARROLL, C. B. COLLINS and I. I. POPESCU,
{\it Hyperfine Interactions}, accepted for publication (1996).
\end{thebibliography}
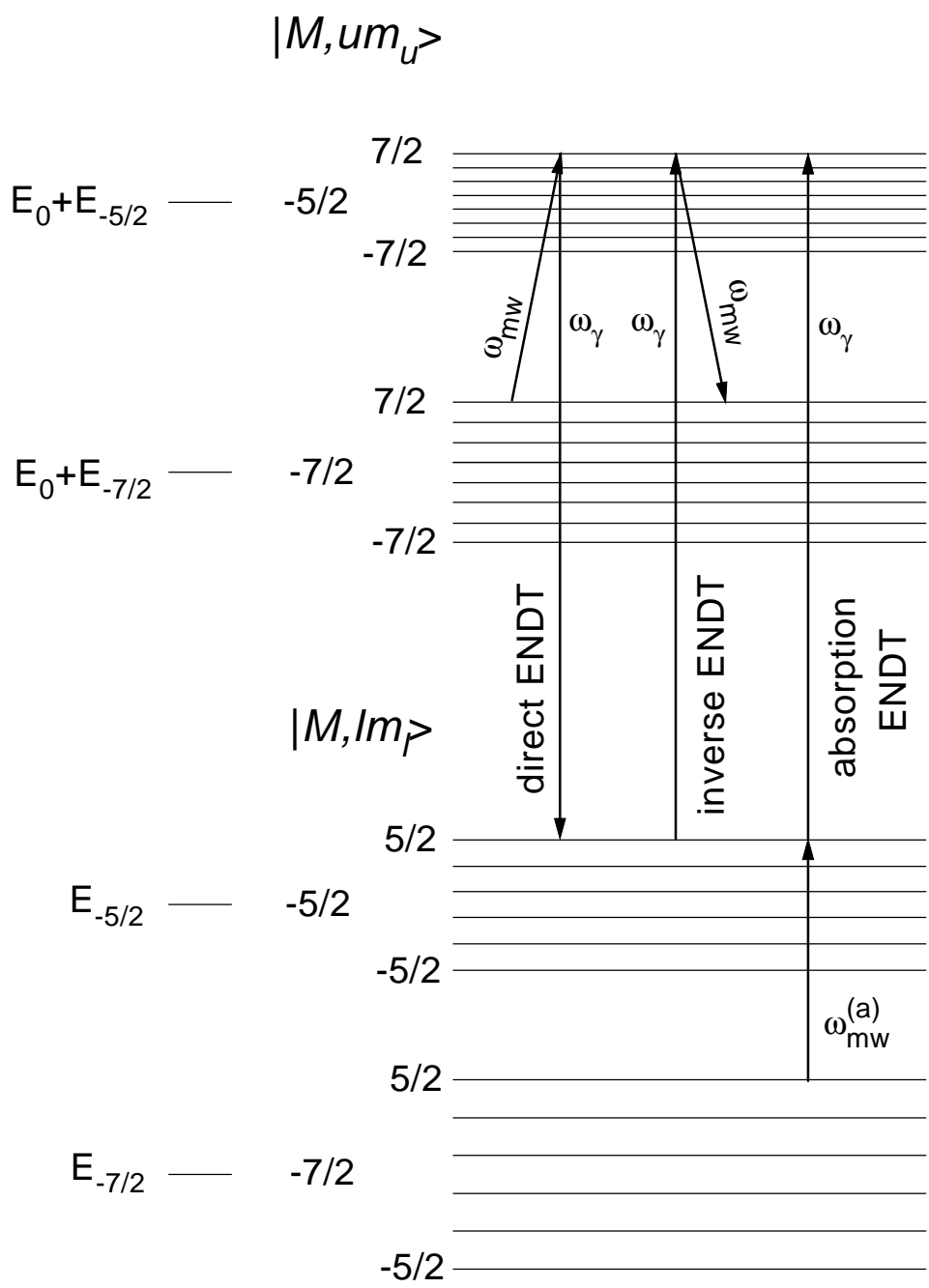
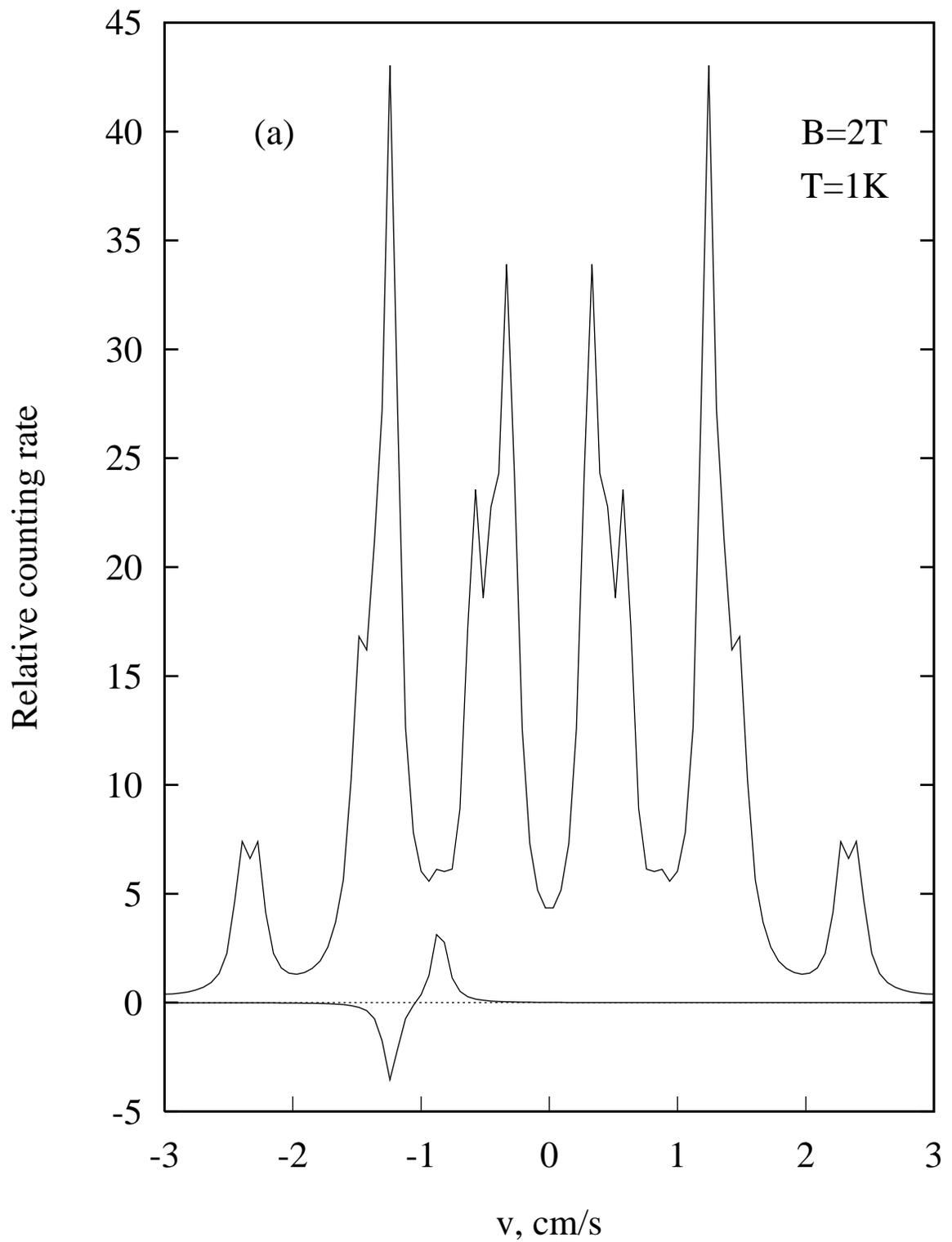
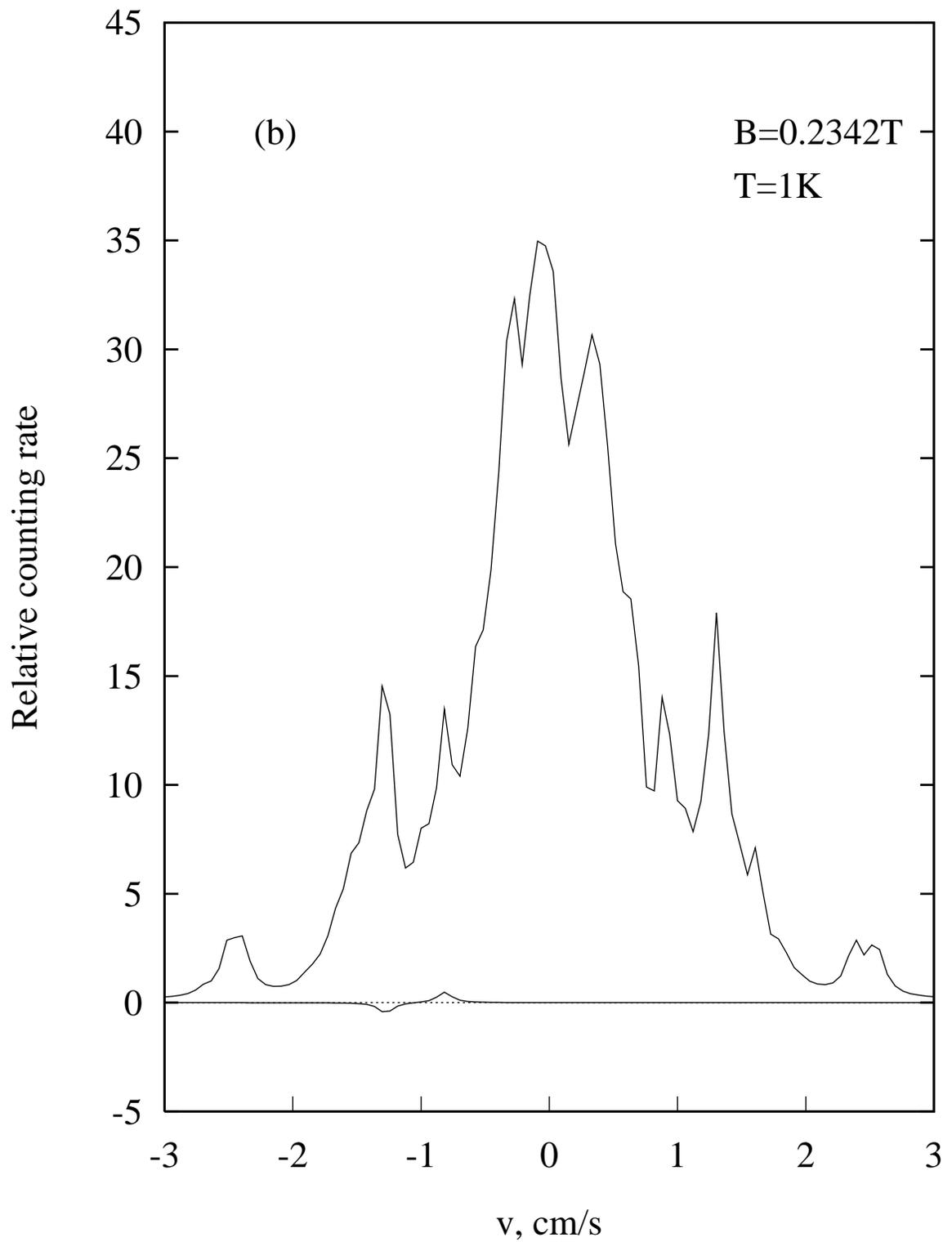
\end{document}